# Nanosized Monoatomic Palladium Metallic Glass

**Authors:** Dong Sheng He[1]†, Yi Huang[1]†, Benjamin D. Myers[2], Dieter Isheim[2], Xinyu Fan[1], Yunsheng Deng[1], Lin Xie[1], Shaobo Han[3], Yang Qiu[1], Li Huang[1], Vinayak P. Dravid[2], Jiaqing He[1]*

**Affiliations:**

[1] Pico Center and Department of Physics, Southern University of Science and Technology, Shenzhen, 518055, China

[2] Department of Materials Science and Engineering, Northwestern University, Evanston, IL 60208, USA.

[3] Department of Materials Science and Engineering, Southern University of Science and Technology, Shenzhen, 518055, China

*Correspondence to: Email: hejq@sustech.edu.cn

†Those author contributed to this work equally

In theory, any monoatomic metal can be put into glass form given sufficient cooling rate [1]. To data, however, only few pure metals in the periodic table—mostly bcc metals, such as tantalum, vanadium, and tungsten, have been reported with amorphous states as metallic glasses experimentally[2-5]. Palladium (Pd), a great hydrogen-affinitive fcc noble metal widely used in catalysis[6], hydrogen storage, sensing, and separation[7], is one of earliest-known major element that composes bulk metallic glass alloys[8], but itself has never been reported with the capability of forming metallic glass experimentally. Here we show, by reducing the size of the system down to nanoscale, monoatomic Pd metallic glass is producible even with very slow cooling rate. The formation process, which was directly captured by *in-situ* transmission electron microscopy, reveals that the Pd metallic glass nanoparticles was resulted from the leaching of silicon (Si) atoms from the Pd-Si droplet. Further hydrogen absorption experiment showed that in sharp contrast to its crystalline counterpart, this Pd metallic glass expanded little upon hydrogen uptake, exhibiting a great potential application for hydrogen embrittlement avoidance. Our results also provide a new insight into the formation of mono-atomic metallic glass at nanoscale.

Metallic glass, or amorphous metal, is a class of peculiar solid materials, where the constituent atoms are organized in a disordered way[9]. It is distinguished from crystalline metals in the regards that they lack translational and rotational symmetry. Thus, the atomic arrangement does not have long-range order. As a result of this disordered atomic structure, metallic glass does not have grain boundary or crystallographic defects, such as dislocations, that are usually the vulnerable part of crystalline materials. Consequently, metallic glass has higher strength, elasticity, and better resistance to wear and corrosion than the crystalline materials[10-12]. Since an amorphous state is not the thermodynamically lowest energy structure, the formation of metallic glass from liquids results from competition between solidification and crystallization. In most cases, a metallic glass is composed of two or more elements, which increases the complexity of the system and thus slows crystallization[13]. However, for the preparation of mono-element metallic glass systems—the jewels in the crown, as it were, crystallization happens so fast that harsh conditions, e.g., ultrahigh cooling rates[2-4] or ultrahigh pressures[5], are believed to be indispensable. To date, only few pure metals—mostly bcc metals, such as tantalum, vanadium, and tungsten, have been reported with amorphous states as metallic glasses[2-5]. Pd is an important noble metal widely used in catalysis[6], hydrogen storage, sensing, and separation[7]. Searching for the amorphous state of Pd will provide a novel freedom to potentially promote these applications. In the liquid state, Pd-based alloys decompose chemically (phase separation) at the bulk size[14-15]. Reducing the size of the initial liquid droplet shortens the diffusion distance and may lead to a complete phase separation in a binary liquid system. In addition, the structure of nanosized particle is more likely influenced or controlled by the substrate than the bulk size. Extending this idea, it should be possible to obtain pure Pd metallic glass at a small size scale.

**Preparation of Pd metallic glass nanoprticles**

In this work, we describe a novel approach for synthesizing pure Pd metallic glass (*a*-Pd) nanoparticles that was facilitated by a chemical reaction between Pd and a silicon nitride (SiNx) support. Fig. 1A shows the schematics of the experiment that produced the *a*-Pd nanoparticles. In a typical experiment, crystalline Pd (*c*-Pd) nanoparticles (Fig. 1B) were dropped on a pre-calibrated heating chip covered with an amorphous 15-nm-thick silicon nitride membrane (Extended Data Fig.1)[16]. The heating chip was mounted on a TEM heating holder. Subsequently, the nanoparticles were heated to and stabilized at 1073 K for approximately 30 minutes in a TEM column with a vacuum environment. Finally, the chip was cooled to room temperature at a controlled cooling rate of 1000 K/s.

The products had a lateral size ranging from a few to hundreds of nanometers, with more than 90% having an amorphous structure (Fig.1C and more evidence in Extended Data Fig.2). The cooling rate had little impact on the productivity of the amorphous nanoparticles (Extended Data Fig.3). A high resolution transmission electron microscopy (TEM) image of the product (Fig. 1E) shows a 'salt and pepper' contrast, similar to the amorphous substrate rather than the lattice fringes. This demonstrates the disorder of the atoms within the product[3]. The electron diffraction patterns of the initial poly-crystalline Pd nanoparticles and the $a$-Pd nanoparticles before and after heat treatment are displayed together for comparison (Fig. 1D). The pattern for the $a$-Pd nanoparticles is more diffusive than the poly-crystalline Pd nanoparticles. Based on the electron diffractions, the radial distribution functions, i.e. the distributions of the inter-atom distances, were calculated (Extended Data Fig. 4-5). The radial distribution function of the amorphous nanoparticles is broader than the crystalline ones. The averaged distance between the $1^{st}$ neighbor atoms in the $a$-Pd nanoparticles is 2% larger than that of the $c$-Pd nanoparticles, which indicates a less dense packing of the atom in the $a$-Pd nanoparticles. Atomic force microscopy measurements show that the morphology of the nanoparticle was flat on the surface (Extended Data Fig. 6), indicating it had experienced a chemical reaction with the substrate during heating. The $a$-Pd nanoparticles were very stable and remained in an amorphous state for more than three months under ambient conditions (Extended Data Fig. 7).

**Compositional characterization of Pd metallic glass nanoparticles**

The composition of the $a$-Pd nanoparticles was then studied using energy dispersive X-ray spectroscopy (EDS), electron energy loss spectroscopy (EELS) and atom probe tomography (APT). Fig. 2A shows the EDS elemental mapping of an $a$-Pd nanoparticle lying on the surface. High angle annular dark filed-scanning transmission electron microscopy (HAADF-STEM) (Fig. 2A) generates contrast as a result of the difference in atomic number (Z contrast). This generally follows the pattern of Pd EDS signals and confirms the presence of Pd in the nanoparticles. The nitrogen (N) and silicon (Si) have relatively weaker signals at the location of the Pd nanoparticle, implying that the substrate was consumed during nanoparticle formation. The HAADF-STEM image and the EDS mapping are very smooth for the $a$-Pd nanoparticles and do not exhibit the precipitates or sub-networks mentioned in other works[14,15,17-19], further confirming the homogeneity of our nanoparticles. A small but visible concentration of Si at the nanoparticle's edge in the line profile (Fig. 2B) will be discussed later.

Two-dimensional EDS mapping alone cannot directly exclude the possibility that the nanoparticles were made of Pd silicide, because of the overlapping of the sample and the Si-containing substrate. A spectral study of the nanoparticle without overlapping of the substrate was possible for the occasional nanoparticle leaning over a hole with a high angle tilt. An EELS line scan was performed in the direction indicated in Fig. 2C and D. Several data points of the Si L-edge were extracted from the full line scan (Fig. 2F; see Extended Data Fig. 8 for full data). There was no Si signal where the electron beam hit the sample but not the substrate (red arrow in Fig. 2E). This shows that this nanoparticle was made of pure Pd rather than Pd silicide (Fig. 2G).

Moreover, to accurately determine the composition of the *a*-Pd nanoparticles, an APT experiment was performed and the results are displayed in Fig. 2G-I. In this experiment, the nanoparticle was sharpened as a sharp tip (Extended Data Fig. 9), and the atoms within the nanoparticle have been extruded in a strong electric field with the help of a laser. The reconstructed 3D atom distribution (Fig.2G-H) shows the nanoparticle was almost made of pure Pd atoms. The line profile (Fig.2I) indicates very little composition variation, confirming the uniformity observed by EDS mapping and EELS line profile. Mass spectroscopy shows the purity of the *a*-Pd nanoparticle was $99.35 \pm 0.23$ at% (Extended Data Fig. 10), which is consistent with the measure-after-etch result using EELS that shows the purity was no less than 99.06% (Extended Data Fig. 11, see Methods for detail).

### *In-situ* investigation of the formation mechanism of Pd metallic glass nanoparticles

While the product we obtained was the same element that we supplied, the formation process was not intuitively straightforward. Here, we used *in-situ* transmission electron microscopy to reveal the details of our vitrification step. Fig. 3A shows TEM sequence images when the sample was heated to 1073 K. Obviously, the initial *c*-Pd nanoparticles were liquefied and spread over the substrate at this temperature (Extended Data Fig. 12 and Movie 1). Pd itself does not melt at this temperature; the nanoparticles must chemically react with the substrate and form a Pd-Si eutectic liquid (with an atomic ratio of approximate 4/1) according to the phase diagram[20] through He Yunhua's reaction (given below):

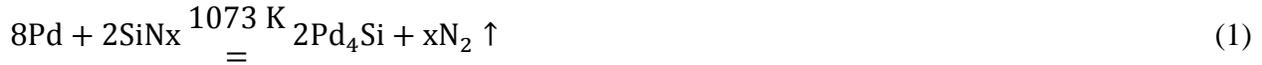

$$8Pd + 2SiNx \xlongequal{1073\ K} 2Pd_4Si + xN_2 \uparrow \tag{1}$$

Our EELS measurement of the liquid droplet confirms the reaction, where the liquid droplets contained Pd and Si at a concentration of 77.73 at% and 22.27%, respectively (Extended Data Fig. 13 and Table 1). The chemical reaction also explains the sacrifice of the substrate observed in Fig. 2.

For the cooling process, both images and EELS were taken to identify the structural and chemical changes. Fig. 3B presents the TEM images of the captured liquid when it was cooled stepwise (full date in Extended Data Fig. 14). Until 623 K, a diffusive ring starts to appear in the FFT pattern, indicating solidification of the droplet. An undercool temperature of at least 400 K in this system represents a very good glass-forming ability. The average cooling rate of this process was between 0.1 K/s to 1 K/s, suggesting that producing *a*-Pd this way does not require harsh cooling conditions (Extended Data Fig. 15). In Fig. 3C, EELS was used to monitor the charge state of Si during the cooling process. Using an amorphous Si spectrum as a reference, it was possible to identify that the Si atoms in the Pd-Si droplet at 1073 K were slightly positively charged (the spectrum is right-shifted). When it reached 1023 K, the Si atoms became neutral and concentrated on the edge (Fig. 2B and Extended Data Fig. 16), showing that they leached out from the liquid and formed amorphous Si (Extended Data Fig. 11). More importantly, the abrupt shift of the spectrum from 1073 K to 1023 K in Fig. 2C indicates a complete leaching-out of Si above 1023 K, consistent with previous compositional measurement of the final metallic glass. Meanwhile, this liquid droplet had not solidified at 1023 K (Fig. 3B), because the surface amorphous Si layer could prevent solidification of the whole droplet[21-22]. This amorphous Si layer also acts as an amorphous template for the solidification of the Pd metallic glass.

**References for main text**

**References for Methods**

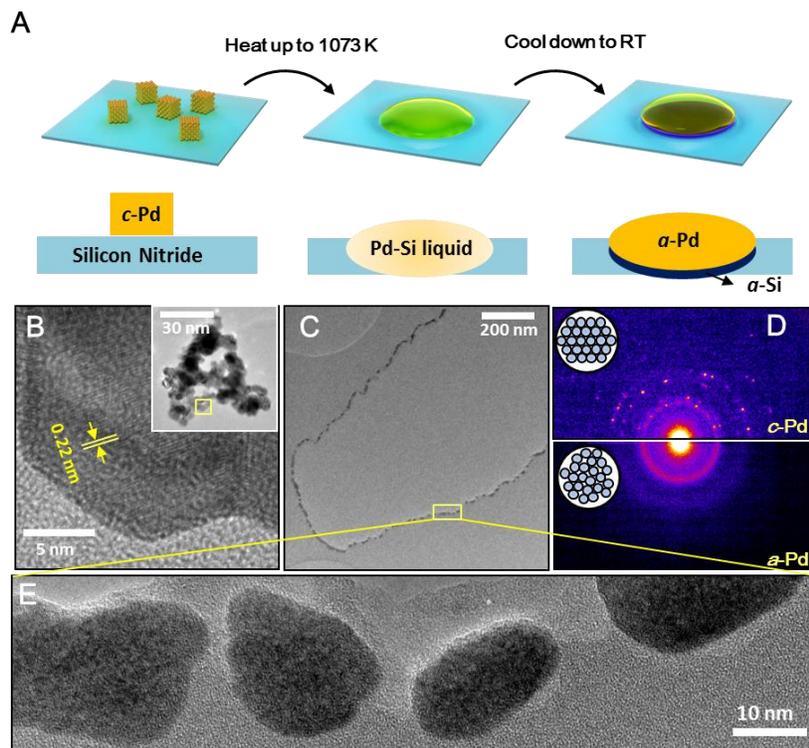

**Fig. 1 Preparation of monoatomic *a*-Pd nanoparticles** (**A**) Schematics of the preparation route. The initial *c*-Pd nanoparticles were subject to heating to 1073 K, and reacting with the silicon nitride substrate to form Pd-Si liquid droplet. During the cooling process, Si atoms leached out and monoatomic *a*-Pd nanoparticles formed. Lower is a sectional view. (**B**) TEM image of the initial *c*-Pd nanoparticles. The periodic lattice fringes, with a spacing of 0.22 nm on the nanoparticle, indicate the crystalline nature of the Pd nanoparticles. (**C**) A low magnification TEM image of monoatomic *a*-Pd nanoparticles after preparation. (**D**) Electron diffraction of initial *c*-Pd (upper) and *a*-Pd nanoparticles (lower). (**E**) High resolution TEM image of *a*-Pd nanoparticles from the area of the yellow box in (**C**).

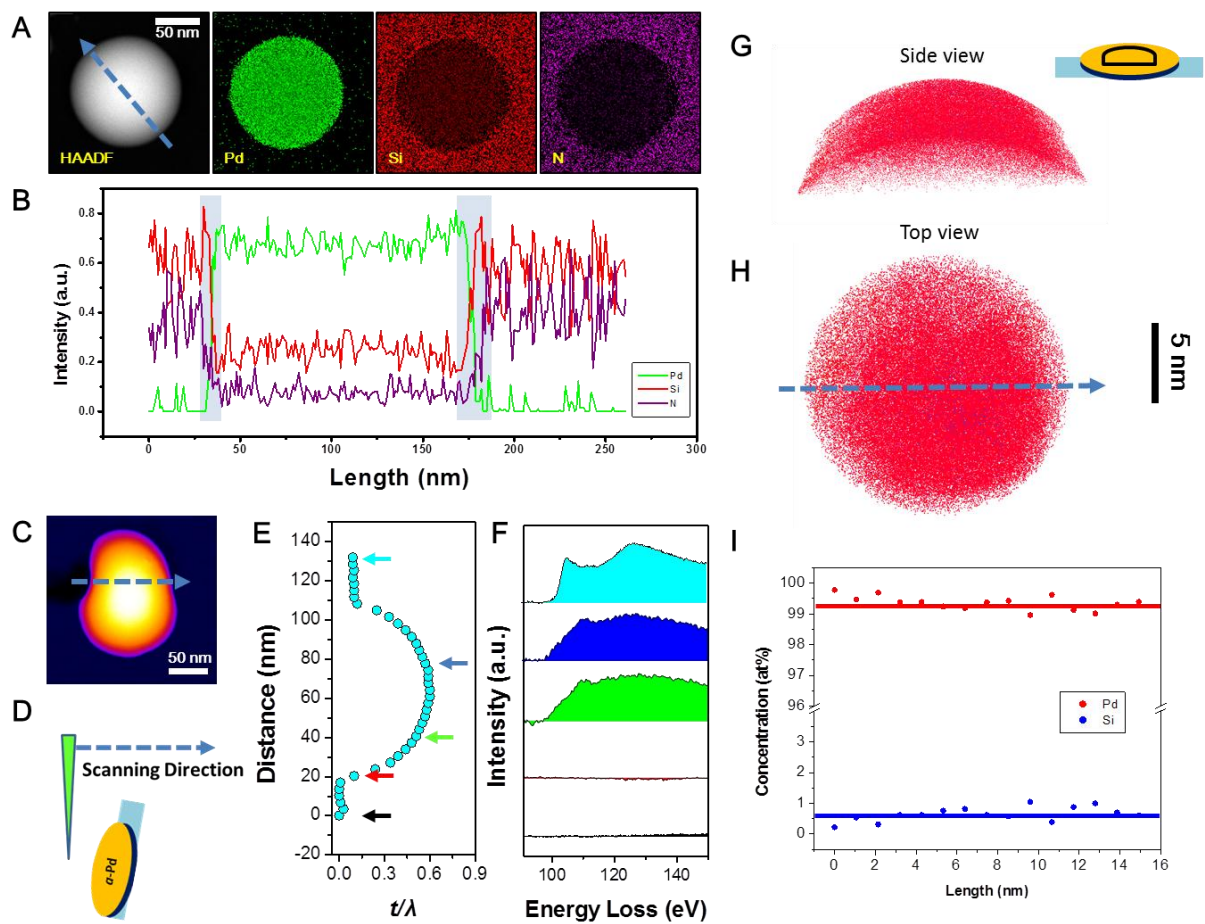

**Fig. 2 Compositional characterization of *a*-Pd nanoparticles**. (**A**) EDS mapping of the *a*-Pd nanoparticle. (**B**) Line profile (direction is indicated in (**A**)) of EDS mapping signals. The transparent blue band highlights the edge of the nanoparticle. (**C**) HAADF-STEM image (false colored) and (**D**) side schematics of a high-angle tilted *a*-Pd nanoparticle. (**E**) Thickness variation along the line profile, which is indicated by the dashed arrows in (**C**) and (**D**). (**F**) EELS at the range of Si L-edge. Sampling positions are indicated by the same color arrow in (**E**). (**G**) Side view and (**H**) top view of the compositional distribution of the *a*-Pd nanoparticle mounted on the APT tip. Pd and Si atoms are represented with red and blue colored dots, respectively. (**I**) Line profile of compositional distribution along the arrow indicated in (**H**). The averaged values are displayed with straight-lines to guide the eyes.

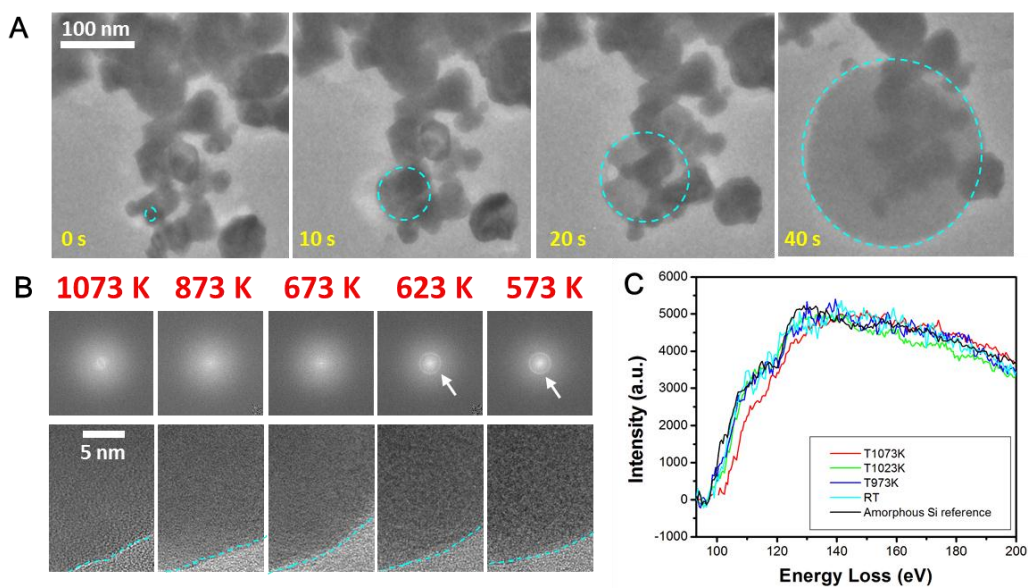

**Fig. 3 Formation of *a*-Pd nanoparticles studied by *in-situ* transmission electron microscopy**. (**A**) Time elapsed TEM images of *c*-Pd nanoparticles on a surface of amorphous silicon nitride at 1073 K. (**B**) TEM images (lower) for the liquid-substrate boundary at different temperatures during cooling. The blue dash lines highlight the nanoparticle–substrate boundary. The left part of the blue dashed line is the nanoparticle. FFTs (upper) were done only for the region of the nanoparticle, and the arrows at 623K and 573K highlight the noncrystalline ring of solid *a*-Pd nanoparticles. (**C**) EELS of a Si L-edge when an Pd-Si droplet was cooled down. Red, green, blue and cyan curves represent the 1073 K, 1023 K, 973 K and room temperature respectively. The black curve was acquired from amorphous Si for comparison.

**Methods**

Preparation of *a*-Pd nanoparticles

For this experiment, the initial *c*-Pd nanoparticles was purchased from Aladdin with purity >99.9%. In fact, the *a*-Pd nanoparticles can also be produced by using different sources of Pd nanoparticles, for example, chemically synthesized[28]. The *c*-Pd nanoparticles were dispersed into ethanol by putting the mixture in the ultrasonic machine for half an hour. Then the solution was dropped on a heating chip covered with amorphous silicon nitride membrane using a pipette. Extended Data Fig. 1 is a typical optical micrograph of the *c*-Pd nanoparticle on the surface of the chip. The chip was purchased from FEI with temperature pre-calibrated (temperature variation <4%). In a typical experiment of generating the *a*-Pd nanoparticles, the chip were heated to 1073 K directly and stabilized for 30 min, and the *a*-Pd nanoparticles can be produced with a controlled cooling rate of 1000K/s. It should be noted that, despite the *a*-Pd nanoparticles were produced in TEM, during the time of preparation, the electron beam was off.

Electron diffraction profiles

All the electron diffraction profiles appear in this paper were obtained by rotationally averaging two dimensional electron diffraction pattern using the Radial Profile plugin in ImageJ. The plugin was developed by Paul Baggethun and was downloaded from: https://imagej.nih.gov/ij/plugins/radial-profile.html.

Radial distribution functions

Radial distribution functions in Extended Data Fig. 5 were calculated from the electron diffraction profiles shown in Extended Data Fig.4 . The data was processed by using the RDFtools plugin in Digital Micrograph[29]. Both crystalline and amorphous data was damped to 15 $nm^{-1}$ using H-Genzel filter. Only Pd was selected for the choice of the element. The density of bulk Pd 12.0 $g/cm^3$ was used. Extended Data Fig. 5 shows the reduced density functions for *c*-Pd nanoparticles and *a*-Pd.

EELS line profile

The data presented in Fig.2F and Extended Data Fig. 8 was deconvoluted by Fourier ratio method. The background was fitted using power law and the background energy window of 91-98 eV was used. The accelerating voltage of electrons was 300 kV. The EEL spectroscope was GIF 966. The camera length was 29.5 mm with spectrometer entrance aperture of 2.5 mm, corresponding to a collection angle of 20 mrad. The convergence angle was 17.9 mrad. The line profile was acquired in a dualEELS mode.

APT experiment and data analysis

The APT sample preparation for the *a*-Pd nanoparticle has been illustrated in Extended Data Fig. 9. In detail, the *a*-Pd nanoparticle was first prepared and confirmed to be amorphous structure in TEM. Then the specific nanoparticle was identified in FIB, transferred and sharpened on the top of an APT sample post.

The APT experiment was carried out in the laser mode with a wavelength of 343 nm. The specimen was cooled down to the temperature of 30 K. The pulse energy was 20 pJ with a rate of 250 kHz. The detection rate was set to 0.10 - 0.50 %.

In the mass spectrum (Extended Data Fig. 10 and Extended Data Table 2), it is visible that minor ions (such as Ga, C, $Pd_3C$) were detected. Comparing with the composition measured within TEM (Extended Data Fig. 13), where the nanoparticle only contained Si and Pd, it is concluded that other ions/atoms are contaminations. In fact, there are three major sources of the contaminations. First, FIB sample preparation can induced Ga, C and possible Si (the substrate contained Si) implanting. Second, Oxygen and the higher-order $O_x$ oxygen cluster ions were likely caused by surface adsorption of water and oxygen, which was not only limited to exposure to atmosphere during sample transfer from TEM to FIB, but this will happen over time even in a ultra-high vacuum. Water and oxygen molecules that are only loosely physisorbed or weakly chemisorbed on the surface can migrate on the surface during atom-probe analysis. Third, the formation of cluster ions was related to the field-evaporation and was usually more pronounced for laser-assisted field evaporation where the tip heats up by ~50-100K. In addition, the catalytic properties of Pd may favor the formation of specific molecular species, here $O_x$, PdO, and $Pd_3C$, in the presence of O (or water) and C from surface adsorption and carbon implantation, respectively. Nevertheless, it is obvious that the amount of Si atoms are even less than the contaminations .When calculating the purity of the nanoparticle, only Si and Pd are taken into account. The standard deviation of the purity along the line profile was used to determine the errorbar.

Measuring the purity of *a*-Pd nanoparticles using EELS (measure-after-etch)

To confirm the purity of the *a*-Pd nanoparticle, we used EELS to make another measurement (Extended Data Fig. 11). In detail, firstly we did the EELS mapping of Si L-edge over the *a*-Pd nanoparticle. As the energy loss near edge structure (ELNES) of amorphous silicon, which is as result of leaching, is significantly different from those within silicon nitride substrate, it is possible to quantitatively separate the count that generated by amorphous silicon and silicon nitride using multiple linear least squares (MLLS) fitting in Digtal Micrograph. Then the heating chip was put in concentrated nitride acid for overnight, to selectively remove the Pd. It is obvious in Extended Data Fig. 11D that at the site of the *a*-Pd nanoparticle, some amorphous silicon remains, confirming the leaching of Si. Do the same EELS Si L-edge mapping and measure the contribution of amorphous Si again. We found the difference in the amount of silicon atom is 3.3 at%. This means that the upper bound for residual silicon within the Pd nanoparticles is 0.94 at% (i.e. the purity of Pd is 99.06 at%), considering the possible loss of silicon atoms during etching.

Stepwise cooling curve

In Fig.3B, the sample was cooled stepwisely to get atomic resolution images of the nanoparticle. The cooling curve is recorded and displayed in Extended Data Fig. 15. The imaging was carried out at the temperature plateaus of the cooling curve. The cooling curves for 0.1K/s and 1K/s are also displayed as a reference in Extended Data Fig. 15. It can be seen that the average cooling rate is between 0.1K/s to 1K/s.

Characterization instruments

The atomic force microscope imaging was done by using the Asylum Research microscope (MFP-3D-Stand Alone) in AC mode. The electron diffractions in Fig.1 and Fig.4, *in-situ* TEM experiments including the EELS data in Fig.3 and hydrogen absorption data in Fig.4 were obtained in FEI Titan environmental transmission electron microscope equipped with Ceta2 camera and Gatan Enfinum 976. The EDS mapping and EELS line profiles were acquired using double-Cs corrected Titan Themis with Super-X EDS detector and Gatan GIF 966. The heating holder used in this experiment was FEI NanoEx-i/v holder. The cooling holder used in this work was Gatan 636 cooling holder. The FIB used in transferring the sample from the heating chip to the FIB grid was FEI Helios 600i. The APT analyses were performed on Cameca LEAP-4000X Si, Madison.

**Acknowledgments** This work was supported by the National Natural Science Foundation of China (51602143, 51702150, 11874194,11774142), Shenzhen Peacock Plan team (KQTD2016022619565991), Natural Science Foundation of Guangdong Province (grant no. 2015A030308001), and the leading talents of Guangdong Province Program (grant no. 00201517). The authors thank Prof. Cai-Zhuang Wang, Prof. Meng Gu, Prof. Zengbao Jiao, Dr. Junhua Luan, Dr Song Liu, Dr. Minghui Wu, Mr. Wei Li and Mr. Yunhua He for helpful discussions. The authors are grateful for the Pico Center at SUSTech core research facilities. The computing time was supported by the Center for Computational Science and Engineering of Southern University of Science and Technology.

**Author contributions** D.S.H. discovered the formation of Pd metallic glass film on silicon nitride membrane by accident. D.S.H and Y.H. carried out most of the characterization of the sample and the *in situ* TEM experiments. D.I, B.M. and D.V.P. did the APT experiment. X.F.and L.H. did the theoretic calculations. L.X. explained the EEL spectra in this work. Y.D. did the AFM experiment. S.H. and Y.Q. did the FIB experiment. J.Q.H. supervised the project and provided critical input. All authors participated in discussion, explained the phenomena in this work, and co-edited the manuscript.

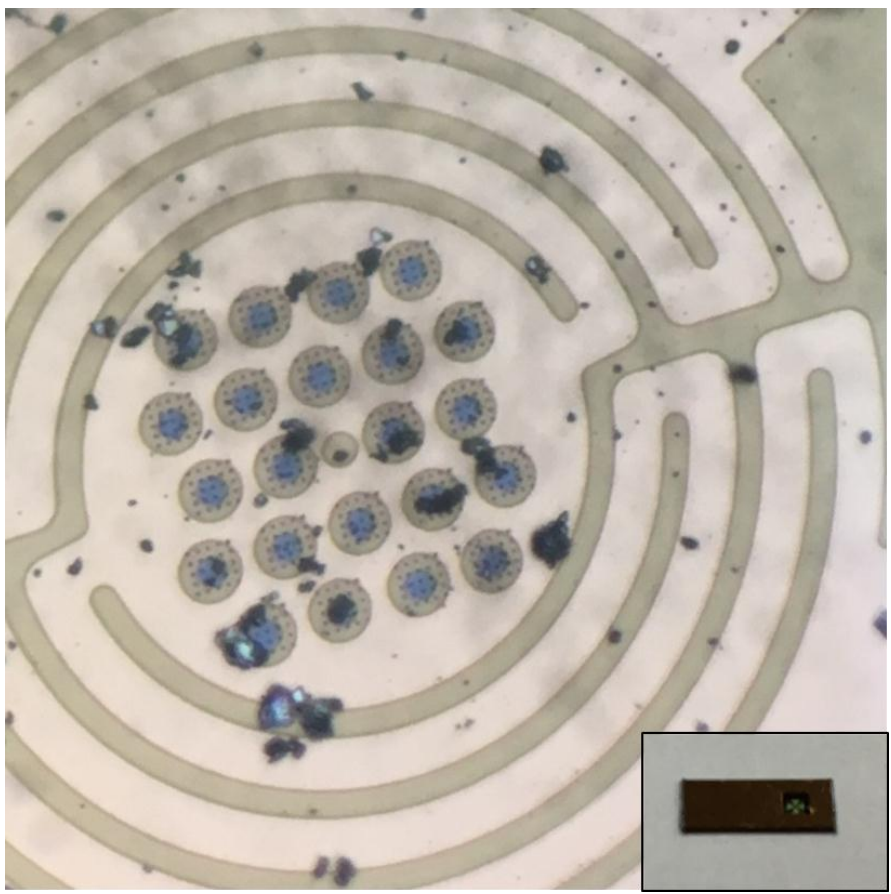

**Extended Data Fig. 1**
Optical micrograph of the initial $c$-Pd nanoparticle deposited on the surface of a heating chip. The inset shows an overview of the heating chip.

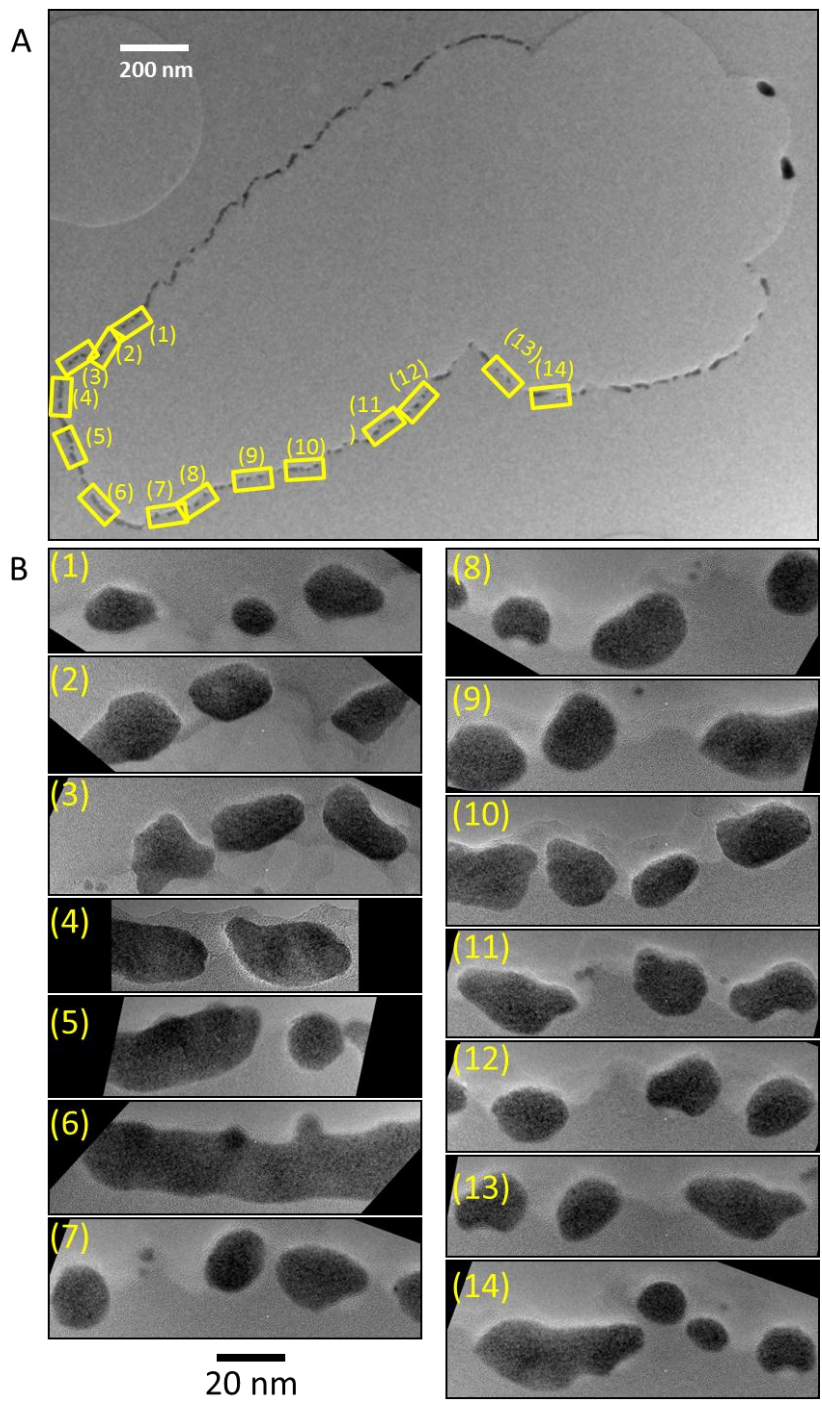

**Extended Data Fig. 2**

(**A**) A low magnification image of *a*-Pd nanoparticles around a local area. (**B**) High resolution TEM images of the *a*-Pd nanoparticles at the areas with the same numbered box in (**A**). It shows that all of the nanoparticles appear to be amorphous, with size ranging from few to tens of nanometers.

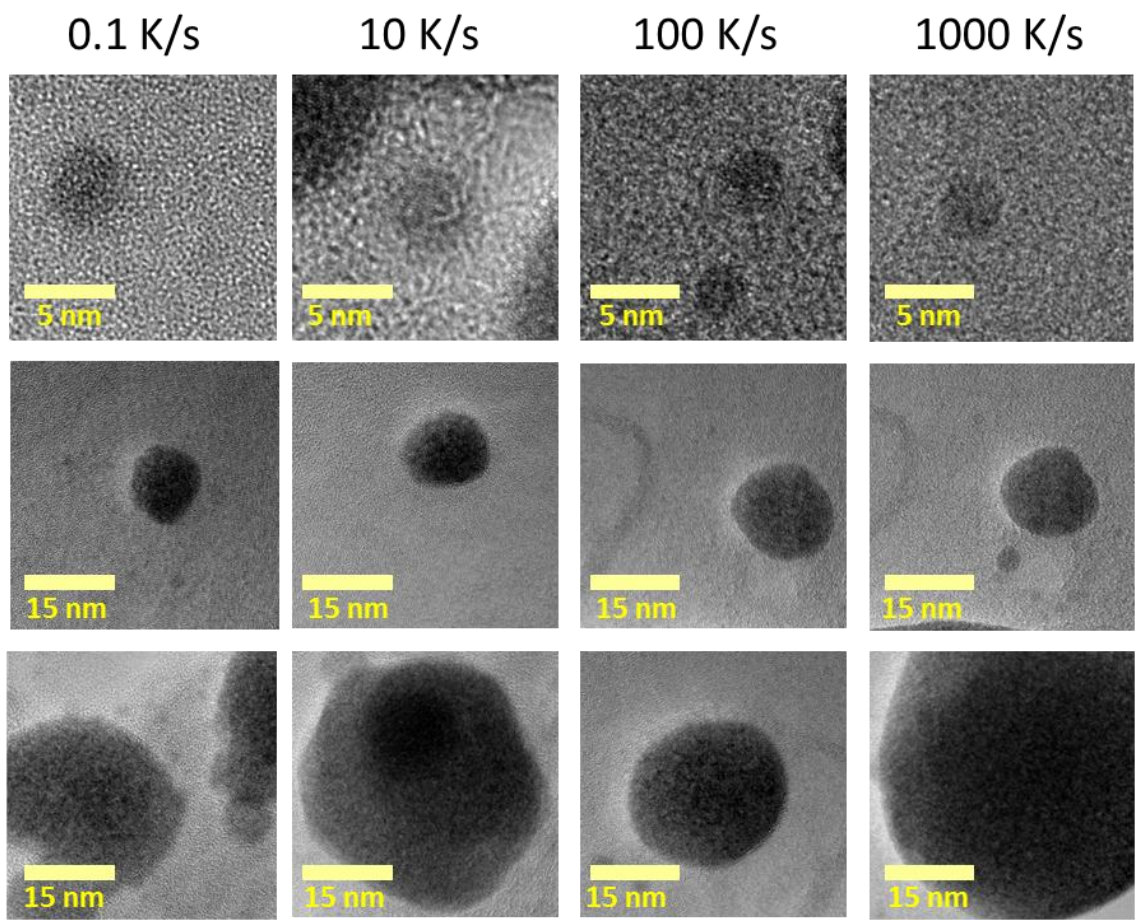

**Extended Data Fig. 3**

High resolution TEM images of *a*-Pd nanoparticles with various size prepared with cooling rate ranging from 0.1 K/s to 1000 K/s (from left to right).

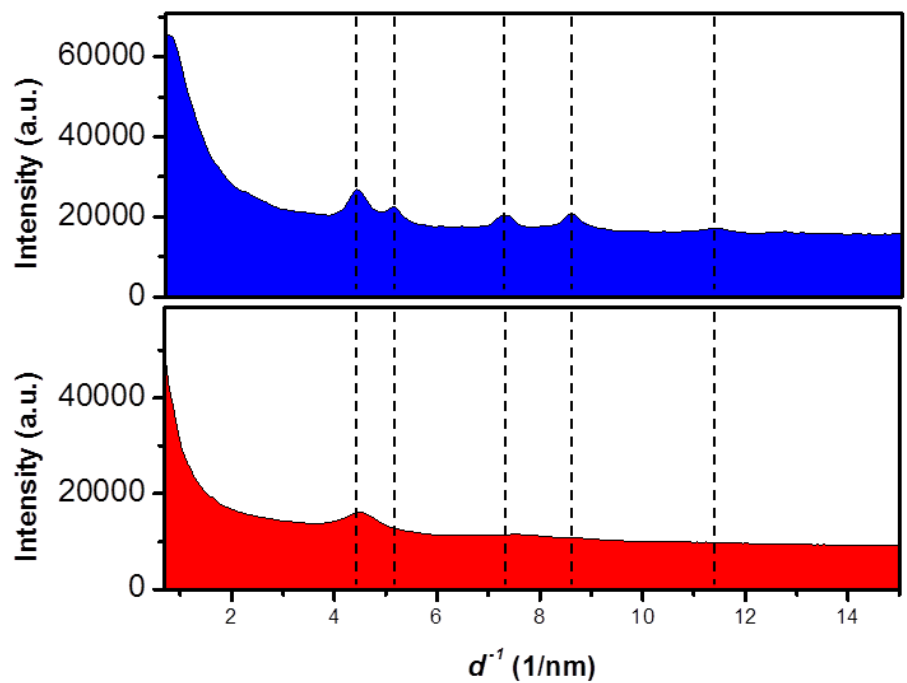

**Extended Data Fig. 4**

Electron diffraction profiles of the initial *c*-Pd nanoparticles (upper) and *a*-Pd nanoparticles (lower) after the heat treatment. The dashed lines are the reference reflections obtained from X-ray diffraction of *c*-Pd.

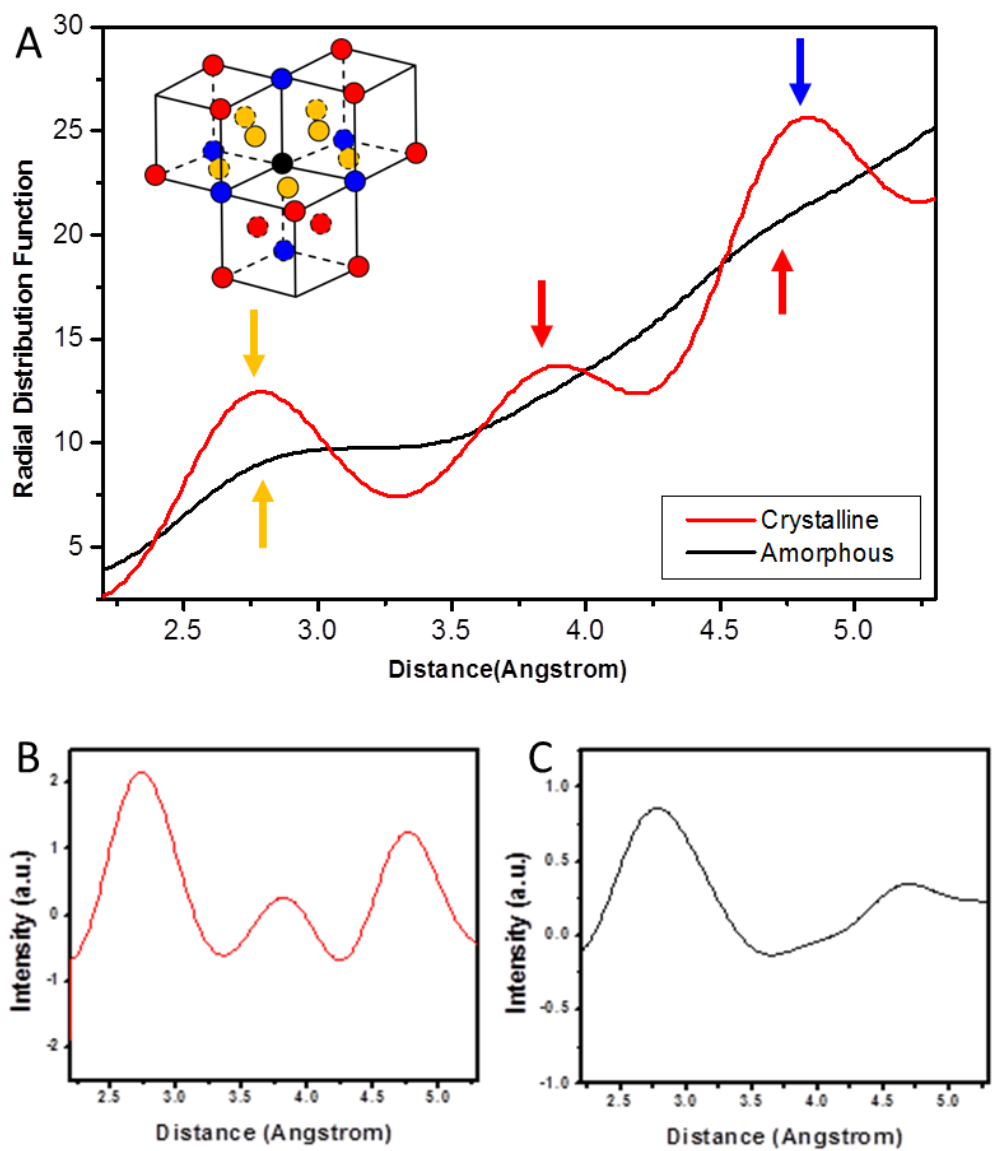

**Extended Data Fig. 5**

(**A**) Radial distribution functions of *c*-Pd and *a*-Pd nanoparticles, calculated from the electron diffraction profiles in Extended Data Fig.4. The inset is the crystalline fcc Pd lattice. Taking the black atom as the center, yellow, red and blue atoms represent the first, second and third nearest neighbors illustrated by the same color arrow for the crystalline radial distribution function. Lower panel shows the corresponding reduced density functions for *c*-Pd (**B**) and *a*-Pd (**C**) nanoparticles.

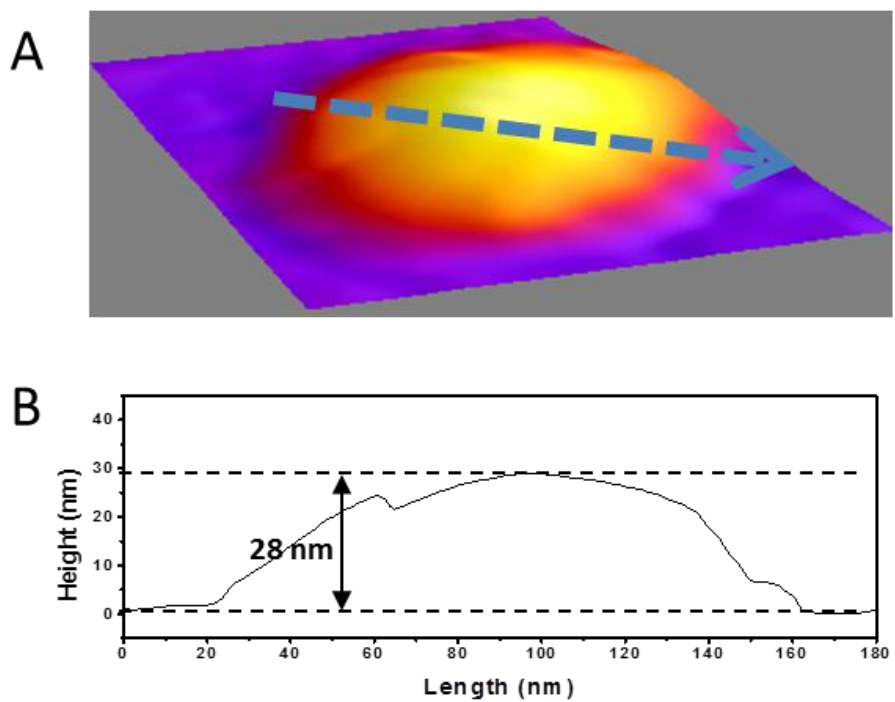

**Extended Data Fig. 6**
 (**A**) Height distribution of an *a*-Pd nanoparticle, which was measured by an atomic force microscope. (**B**) Line profile of the height distribution indicated by the blue dashed line in (**A**).

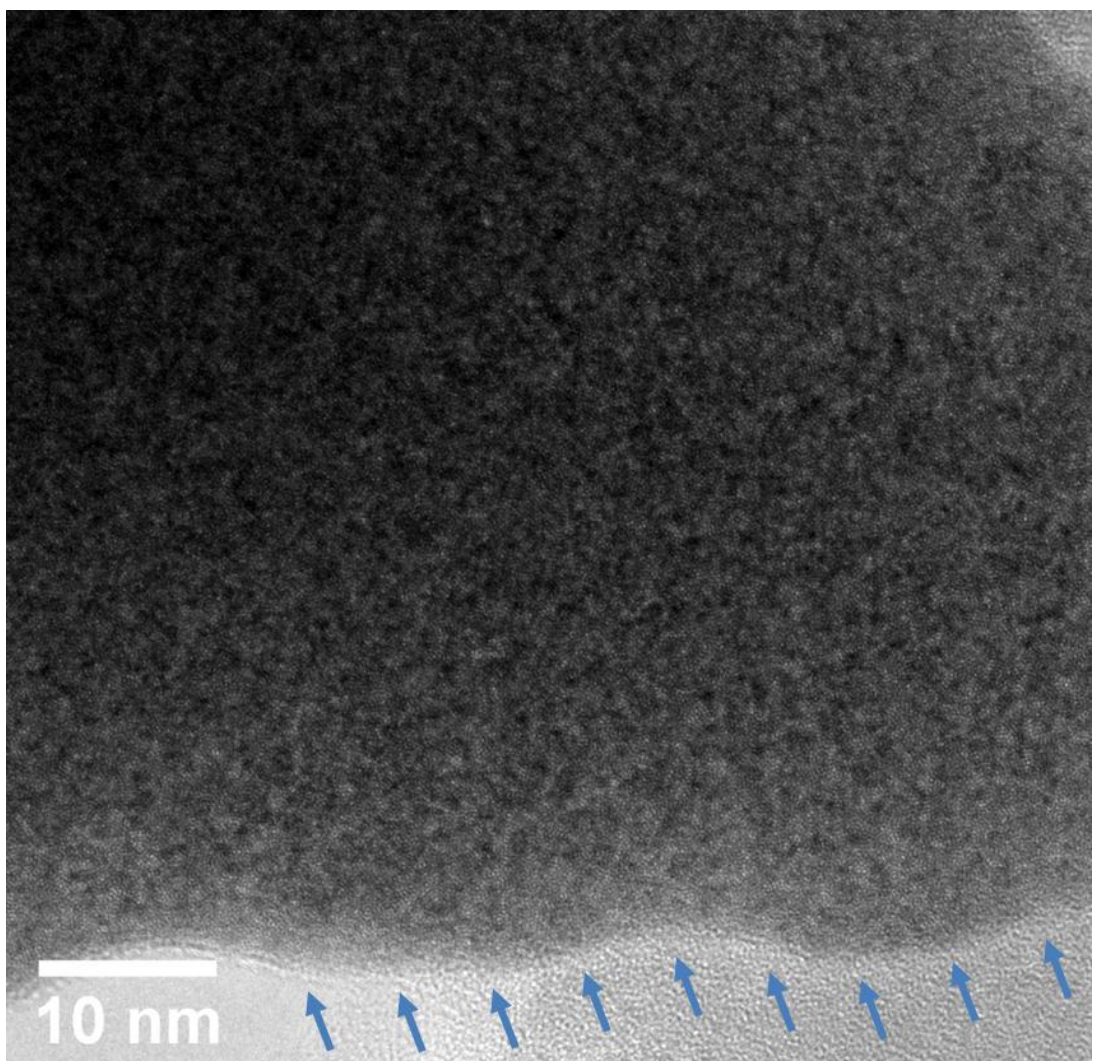

**Extended Data Fig. 7**

High resolution TEM image of an *a*-Pd nanoparticle which has been stored under ambient condition for three months. It is notable that even the surface (indicated by the arrows) did not crystallize.

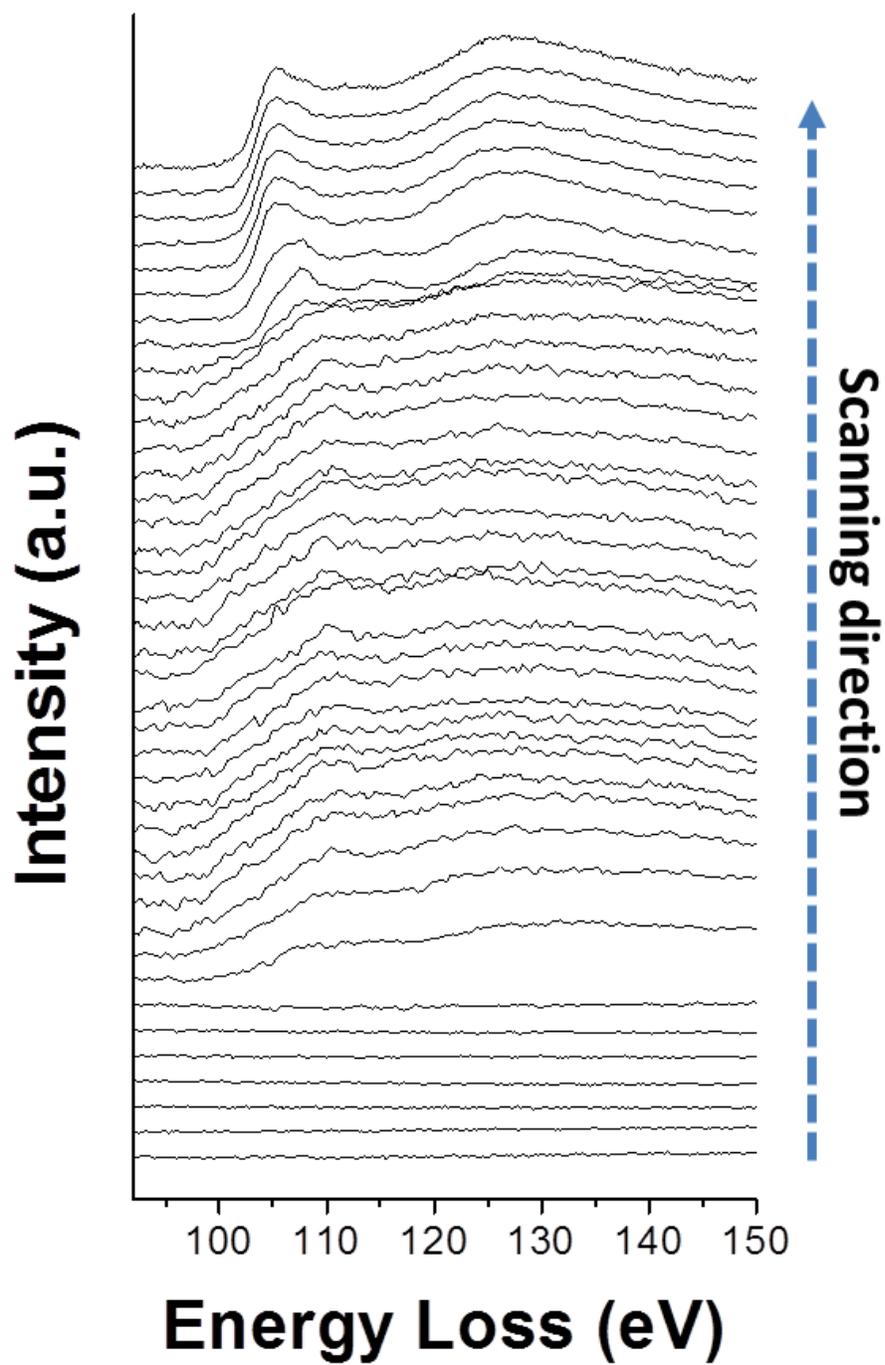

**Extended Data Fig. 8**

Full EELS line profile at the energy range of Si L-edge of **Fig.2F**, they were deconvoluted using Fourier-ratio method and background-subtracted.

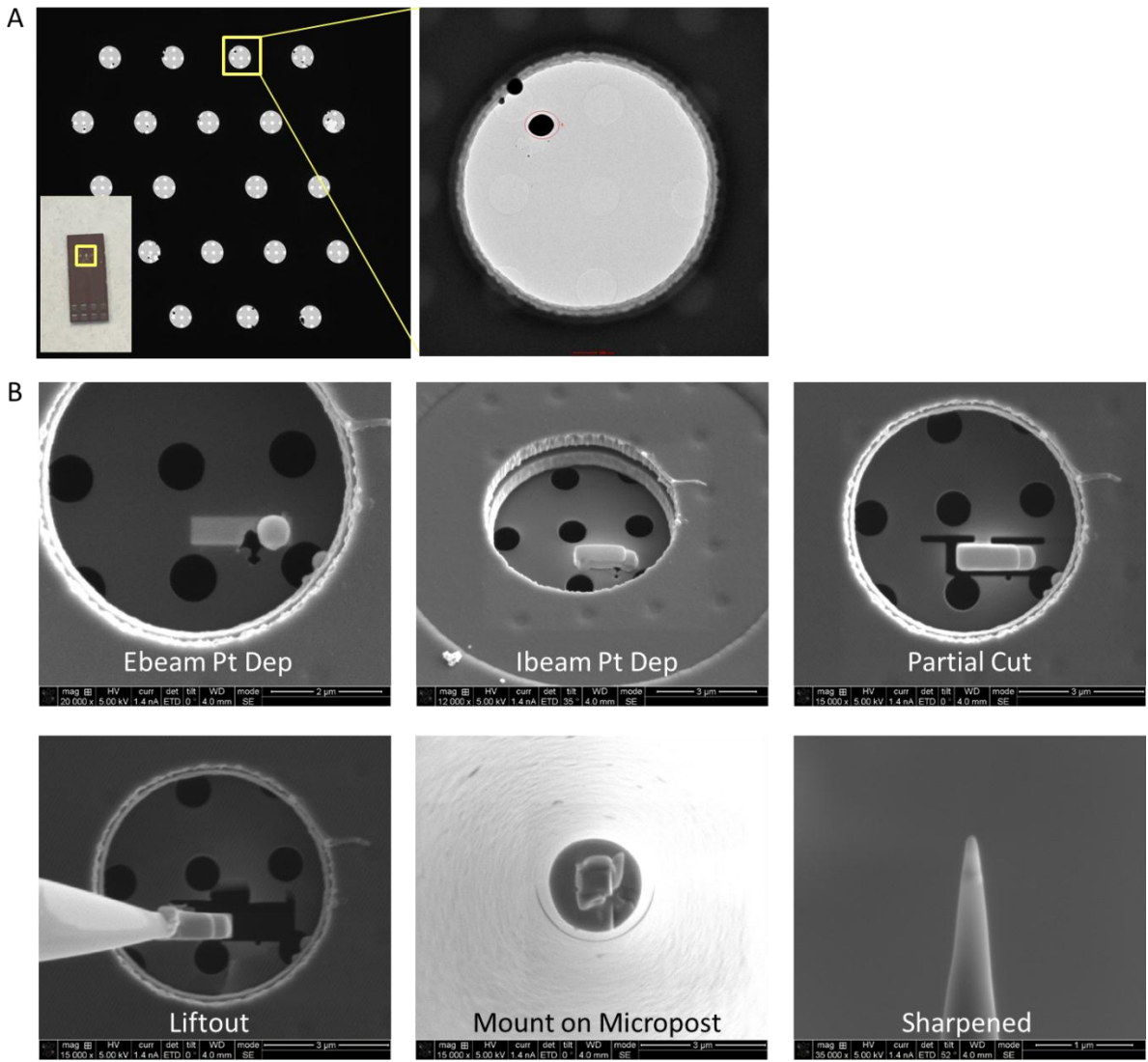

**Extended Data Fig. 9**

Step-by-step illustration of transferring the *a*-Pd nanoparticle on the silicon nitride surface to the APT tip. (A) The *a*-Pd nanoparticle has been identified in the TEM (B) The same nanoparticle has been prepared as an APT sample by Pt deposition, cutting, mounting and sharpening.

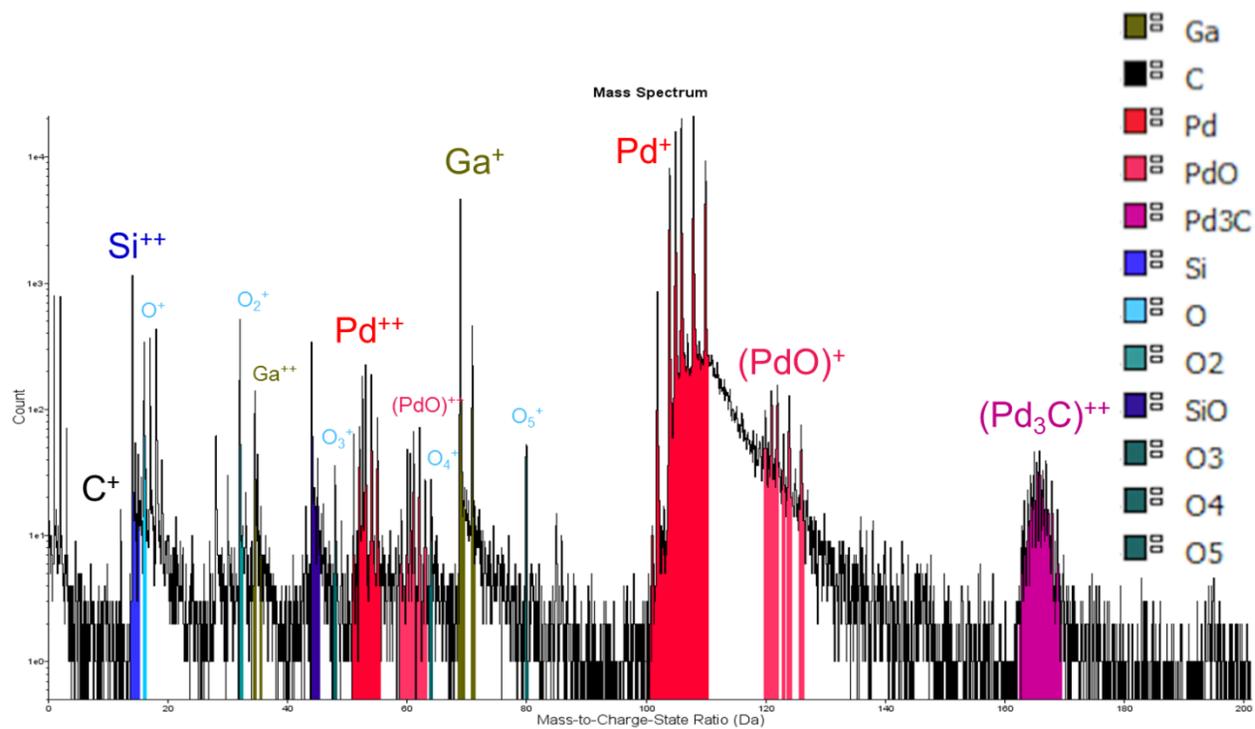

**Extended Data Fig. 10**

Total mass spectrum of the *a*-Pd nanoparticle presented in **Fig.2G-H**, measured by APT. Note the vertical axis is displayed in log-scale.

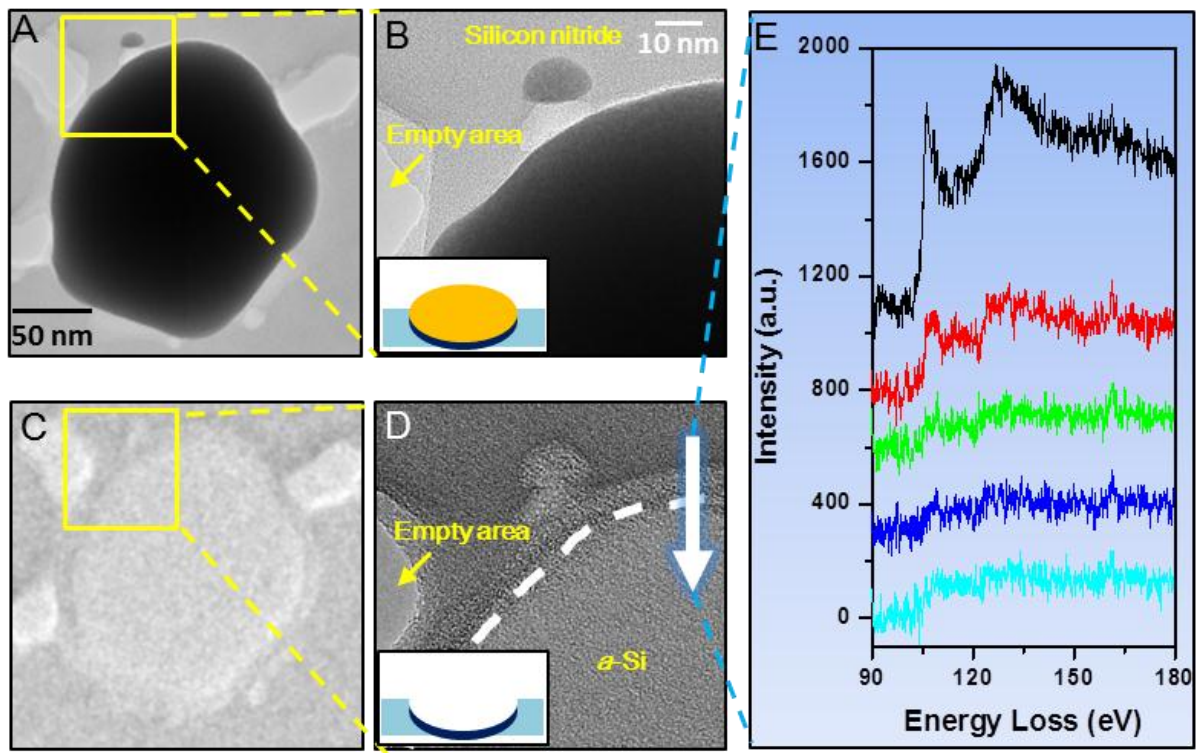

**Extended Data Fig. 11**

(**A**) An *a*-Pd nanoparticle before concentrated HNO$_3$ treatment; (**B**) Magnified area of the yellow box in (**A**). (**C**) The same area of (**A**) after concentrated HNO$_3$ treatment. (**D**) is the same area of (**C**) after concentrated HNO$_3$ treatment. (**E**) EELS line profile with the direction indicated by the arrow in (**D**). In both (**C**) and (**D**), an empty area is displayed to show the contrast of vacuum. (**A**) and (**C**), (**B**) and (**D**) share the same scale bars. The leaching of amorphous silicon (*a*-Si) is confirmed structurally by image in (**D**) (right bottom area) and chemically by EEL spectra (**E**).

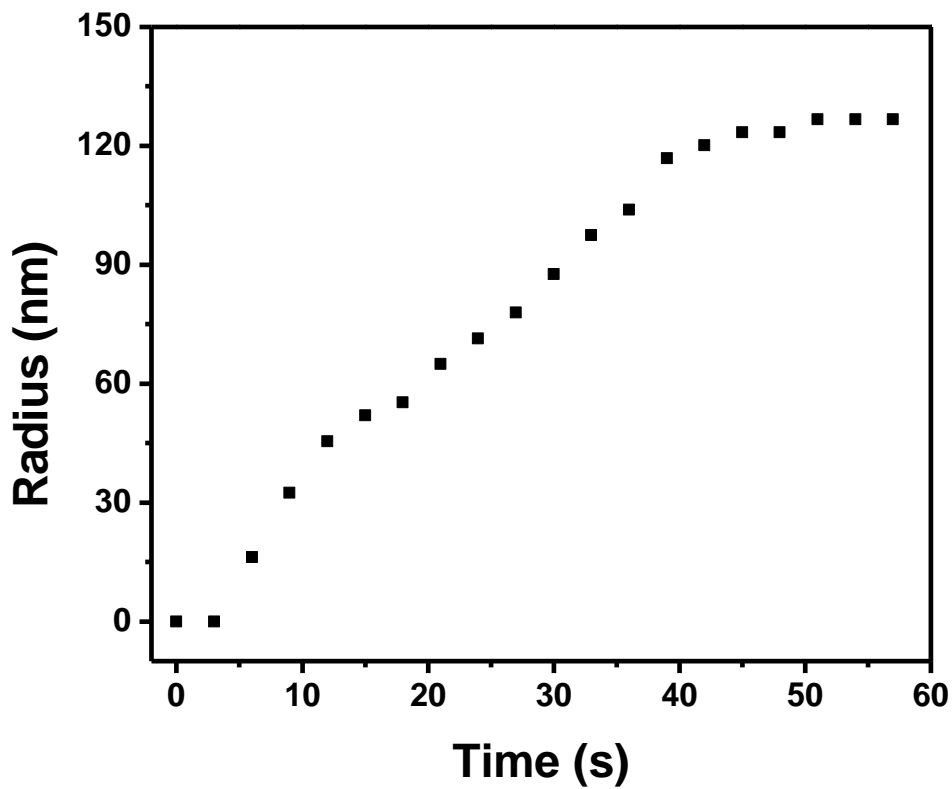

**Extended Data Fig. 12**
The radius of the Pd-Si liquid droplet as a function of time at 1073K. By using a circle to fit the shape of the liquid droplet, it is possible to estimate its size shown in the Movie S1. The radius increases linearly as a function of time. The spreading rate was approximately 3 nm/s.

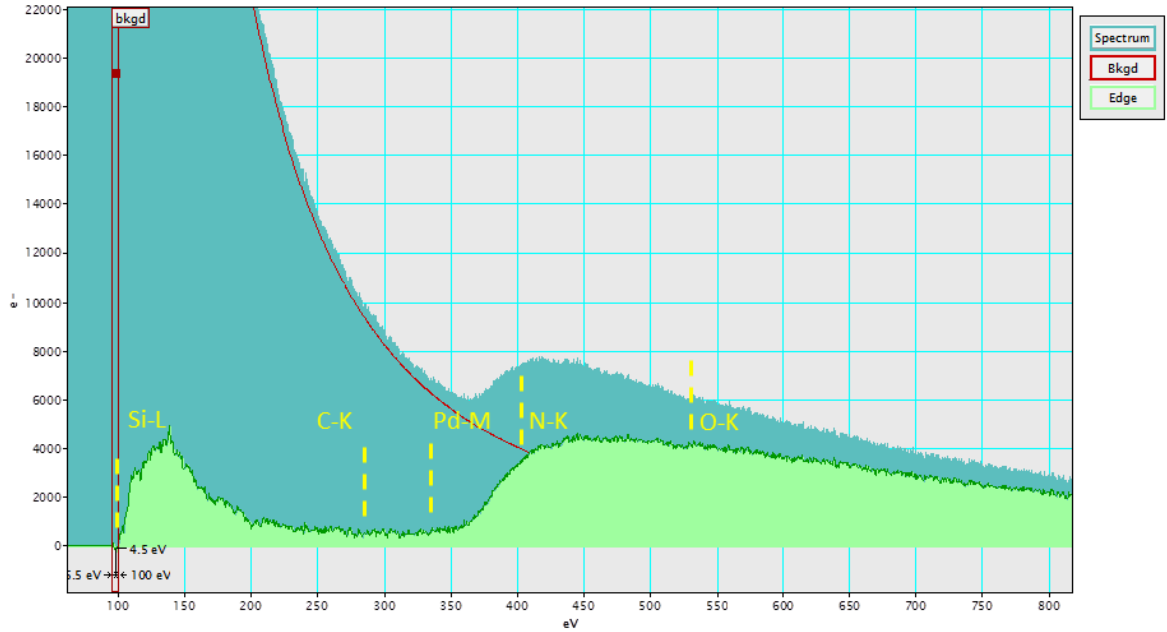

**Extended Data Fig. 13**

The EELS spectrum of the Pd-Si liquid droplet at 1073 K. The edges for common contamination element are labeled for reference (C K-edge: 284 eV, N K-edge: 401 eV, O K-edge:532 eV). It is obvious that only Si L-edge (99 eV) and Pd M-edge (very delayed, 335 eV) is visible. Because the droplet etched the substrate completely (hanging on an empty hole because of the surface tension), there is no nitrogen signal detected here.

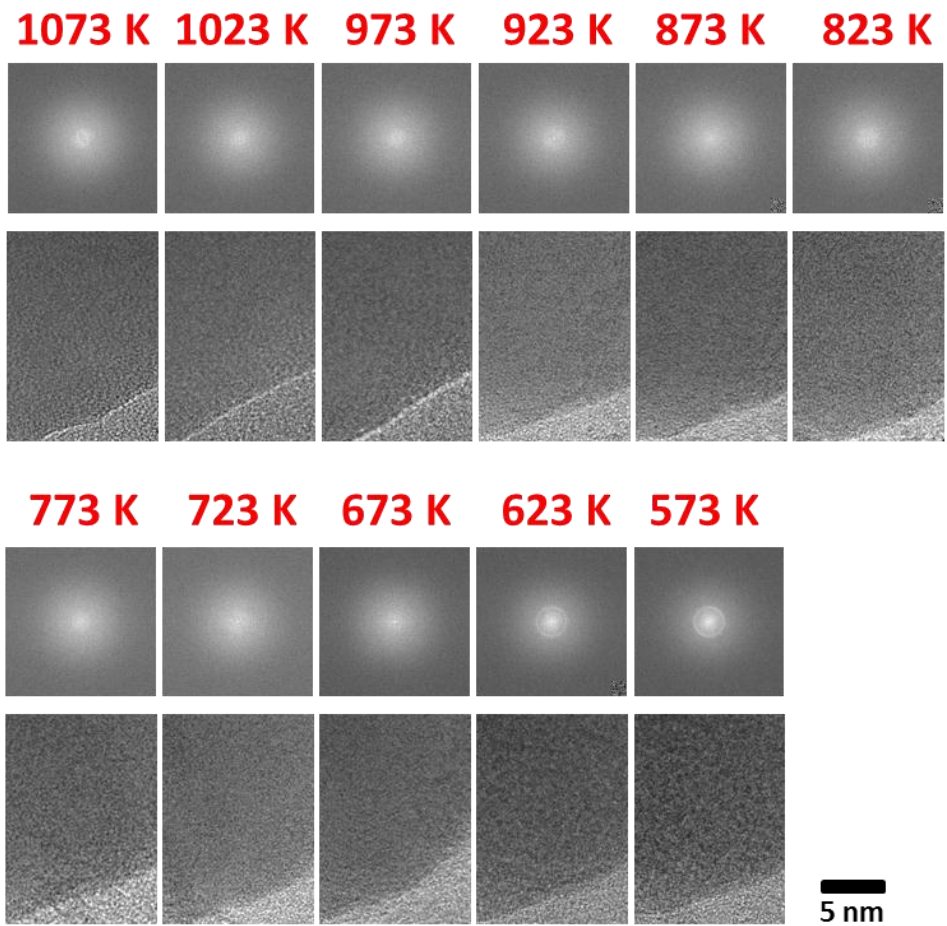

**Extended Data Fig. 14**

Complete image sequence and corresponding FFT of **Fig.3B**, with decreasing 50K per step.

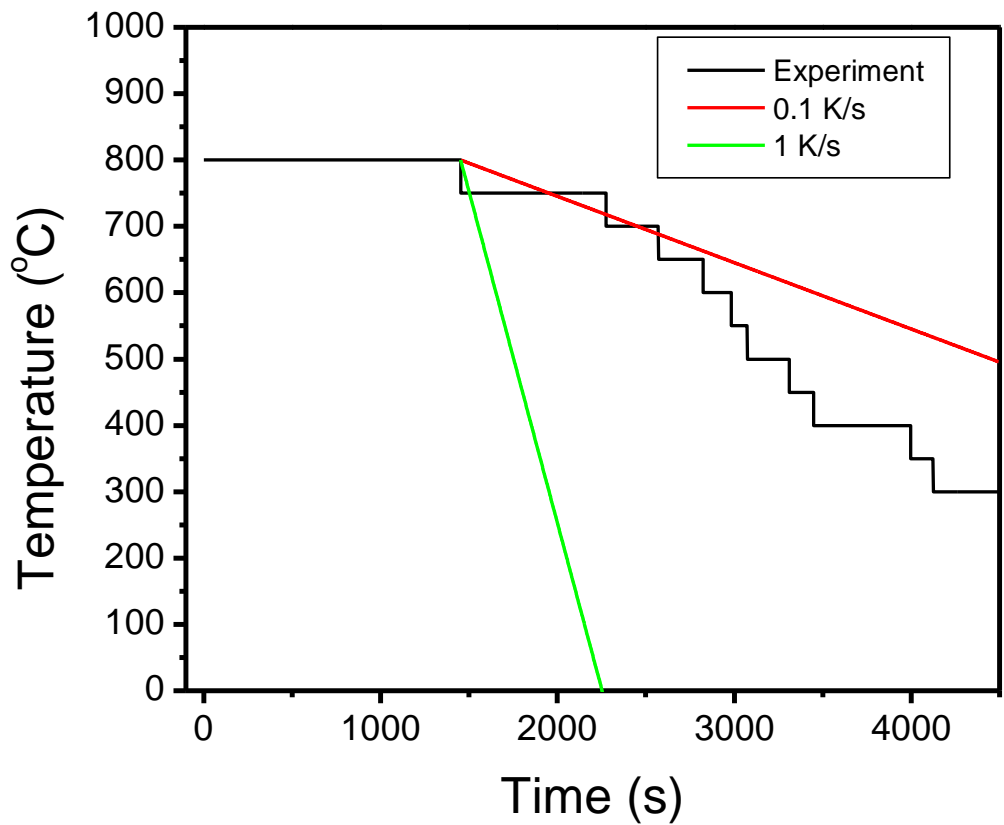

**Extended Data Fig. 15**

Experimental temperature curve during cooling for **Fig.3B**. The cooling curves for 0.1K/s (red) and 1K/s (green) are also displayed as a reference.

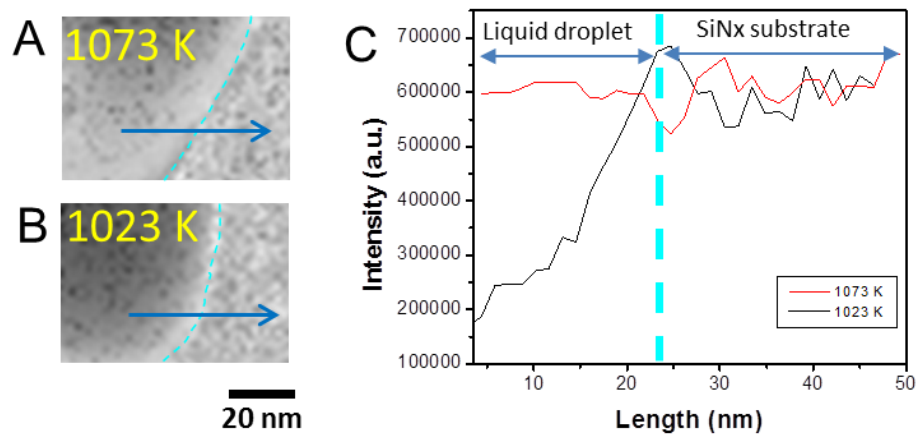

**Extended Data Fig. 16**
EELS mapping of the Pd-Si droplet at (A) 1073 K and (B) 1023 K with integrated Si L-edge energy loss window (98 eV to 198 eV) after background subtraction. (C) Line profiles indicated by the arrows in (A) and (B). Cyan dashed line highlight the boundary of the droplet. The left side of the cyan dashed line is the Pd-Si droplet. It can be seen that the Si signal concentrated from all over the the droplet to the edge when it was cooled from 1073 K to 1023 K.

| Measurement NO. | 1 | 2 | 3 | 4 | 5 | 6 |
|---|---|---|---|---|---|---|
| Pd at% | 77.59 | 75.30 | 76.75 | 83.88 | 78.59 | 74.29 |
| Si at% | 22.41 | 24.70 | 23.25 | 16.12 | 21.41 | 25.71 |

**Extended Data Table 1**

Quantitative measurement of the composition of Pd and Si in the liquid droplet at 1073K .

| Ion type | Ranged% | Atom Type | Ranged% |
|---|---|---|---|
| Pd | 94.1836% | Pd | 94.1247% |
| PdO | 0.6702% | Si | 0.6158% |
| $Pd_3C$ | 0.5015% | Ga | 3.8157% |
| Si | 0.5087% | C | 0.5066% |
| SiO | 0.1217% | O | 0.9372% |
| Ga | 3.9062% | | |
| C | 0.0171% | | |
| O | 0.0468% | | |
| $O_2$ | 0.0308% | | |
| $O_3$ | 0.0021% | | |
| $O_4$ | 0.0035% | | |
| $O_5$ | 0.0077% | | |

**Extended Data Table 2**

The ions were detected by APT experiment and the corresponding percentage within the range of measurement (left two columns). The right two columns show the atomic ratio after decomposing the ions. Both of the data has been calculated after background subtraction.